# Two-dimensional multiferroic NbPc COF with strong magnetoelectric coupling and room-temperature ferroelectricity

Wei Li[1], Dongyang Zhu[2], Shuai Dong[1], and Jun-Jie Zhang[1,3*]

[1] Key Laboratory of Quantum Materials and Devices of Ministry of Education, School of Physics, Southeast University, Nanjing 211189, China

[2] Department of Chemical and Biomolecular Engineering, Rice University, Houston, Texas 77005, United States

[3] Department of Materials Science and NanoEngineering, Rice University, Houston, Texas 77005, United States

Email: junjiezhang@seu.edu.cn

The realization of two-dimensional multiferroics offers significant potential for nanoscale device functionality. However, type-I two-dimensional multiferroics with strong magnetoelectric coupling, enabling electric field control of spin, remain scarce. In this study, using density functional theory and Monte Carlo simulations, we predict that the niobium phthalocyanine covalent organic framework (NbPc COF) monolayer exhibits type-I multiferroic behavior, with a ferroelectric transition occurring above room temperature. Remarkably, the strong magnetoelectric coupling in NbPc COF monolayer arises from the same origin of magnetism and ferroelectricity. Our findings offer flexible pathways for the design and development of organic nanoscale multiferroic devices with broad applications.

**INTRODUCTION**

The coexistence of magnetism and ferroelectricity in materials, known as multiferroics, enables the coupling and control of various ferroic orderings, presenting potential applications in high-performance memory, spintronics, and more [1,2]. However, the post-Moore era introduces challenges in miniaturizing devices using conventional bulk multiferroics [3–7]. Recent theoretical and experimental advances have demonstrated magnetism and ferroelectricity in atomic layers, opening new possibilities beyond the limitations of inactive layer in ferroic film [8–14].

Importantly, a few 2D multiferroics have been recently designed in theory, such as multiferroic van der Waals (vdWs) heterobilayers [15], but their magnetoelectric (ME) coupling is rather weak, primarily due to the independent origins of polarization and magnetism (i.e., type-I multiferroics). Alternatively, type-II multiferroicity was predicted in $Hf_2VC_2F_2$ MXene monolayers and subsequently experimentally observed in $NiI_2$ monolayer [16,17], where ferroelectric polarization is directly induced by frustrated magnetism and can be fully controlled by magnetic fields. Nevertheless, type-I multiferroics with strong ME coupling, allowing electric field control of the spin, are still very promising in the aspects of multiferroic memories. Comparing with inorganic monolayers, 2D vdWs organic systems, particularly covalent organic frameworks (COFs) [18–20], provide a more design flexibility for achieving type-I multiferroicity with strong ME coupling. Additionally, COFs exhibit higher stability compared to organic analogs of metal-organic frameworks [21], and their interlayer interactions are relatively weak, facilitating the exfoliation from bulk to single layers.

In this study, we employed density functional theory (DFT) to design a new type of COF based on niobium phthalocyanine (NbPc) [22]. Our calculations identified the emergence of magnetism and ferroelectricity in this 2D NbPc COF. Using Monte Carlo (MC) simulations, we estimated that the Curie temperature ($T_C$) of the NbPc COF was above room temperature. Importantly, despite the weak magnetic coupling, the strong ME coupling was predicted in 2D NbPc COF, which could be attributed to the same origins of magnetism and ferroelectricity. These findings demonstrate the feasibility and potential of 2D NbPc COF for nanoscale multiferroic devices, making them of great interest to the multiferroic community.

**COMPUTATIONAL METHODS**

The DFT calculations were performed using the plane wave basis Vienna ab initio simulation package (VASP) code [23,24]. The generalized gradient approximation in the Perdew-Burke-Ernzerhof (GGA-PBE) formulation is adopted with a cutoff energy of 600 eV [25]. 2D Brillouin zones are sampled using 2×2 grid k-points in the Monkhorst-Pack scheme, and a vacuum space of ~12 Å is intercalated into interlamination to eliminate the interaction between layers. For the exchange and correlation, the DFT+$U$ approximations are employed to describe the correlation effects on the localized 4d orbital of $Nb^{2+}$ ions, where Hubbard effective potential $U_{eff}$ is considered [26]. The phonon modes are calculated using density functional perturbation theory (DFPT) [27], which is implemented in the VASP code. The evaluation of polarization is conducted using the Berry-phase method [28,29], wherein both the ionic and electronic contributions are taken into account. The nudged elastic band (NEB) method is employed to generate and calculate a series of transition states between the ferroelectric and paraelectric states [30].

The standard Markov chain Monte Carlo (MC) method with the Metropolis algorithm is adopted to investigate the phase transitions [31]. The simulation is performed under periodic boundary conditions with a lattice size of 12×12. The initial $1.5 \times 10^4$ steps are allocated to achieve thermal equilibrium, followed by an additional $5 \times 10^3$ steps for measurements. The acceptance ratios of MC updates are controlled to be ~50% at all simulated temperatures by adjusting the updating windows. The quenching process is utilized in our MC simulations, i.e., gradually cooling the system from a high temperature to the desired temperature. This allows for an effective exploration of the phase space and accurate determination of the phase transitions.

**RESULT AND DISCUSSION**

**A. STRUCTURE**

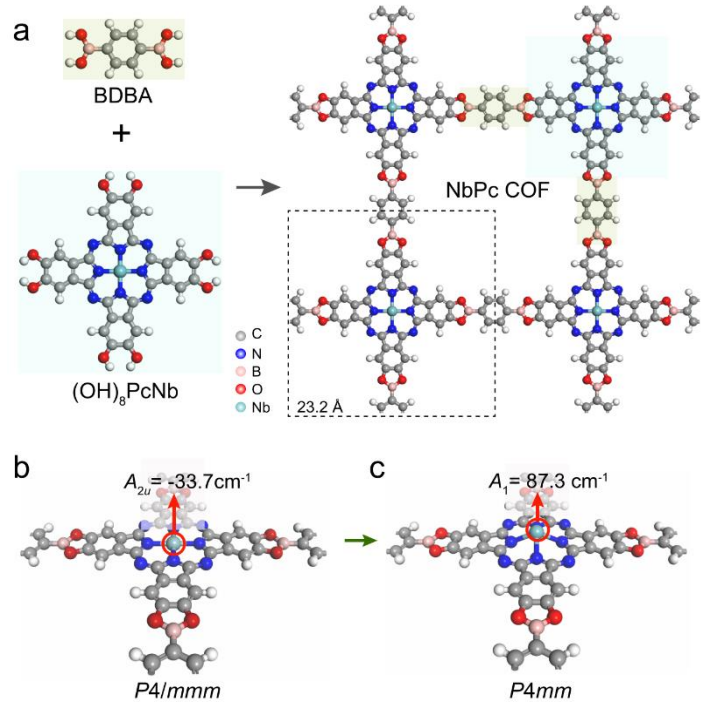

**Fig. 1.** (a) The top view of NbPc COF monolayer that built by BDBA and [(OH)$_8$ PcNb] molecule. The dashed line represents unit cell of NbPc COF. (b) The side view of non-polar phase with P4/mmm (PE), and the sketch of the largest unstable $A_{2u}$ mode. (c) The side view of polar phase with P4mm (FE), and the sketch of the stable $A_1$ mode.

Designing COFs that exhibit the coexistence of ferroelectricity and magnetism is very promising since COFs are composed of diverse organic linkages [18–20], which provide a flexible platform for studying phase transitions and ME coupling in different types of lattice. Motivated by experimentally observed COFs [32], we began by constructing niobium phthalocyanine-based COF (NbPc COF) with nonpolar *P4/mmm* symmetry (paraelectric, PE), as shown in Fig. 1(a).

The NbPc COF is considered as the combination of [(OH)$_8$ PcNb] and 1,4-Benzenediboronic Acid (BDBA) [Fig. 1(a)]. In contrast to the NiPc COF monolayer, the NbPc COF monolayer appears an unexpectedly soften $A_{2u}$ mode at Γ point [Fig S1(a-b) in Supplemental Material (SM)] [33], resulting in an out-of-plane displacement of Nb atoms relative to the atomic plane [Fig. 1(b), left]. Thus, this soften mode breaks the out-of-plane mirror symmetry and yields a new phase of polar *P4mm* symmetry [Fig. 1(b), right]. Now, the polar phase becomes dynamically stable [Fig S1(c-d)] [33], suggesting the existence of ferroelectric transition in NbPc COF monolayer. To evaluate its thermal stability, the binding energy is calculated as $E_p$ = $E$(NbPc COF)- $E$[(OH)$_8$PcNb]-$E$(BDBA)+8$E$(H$_2$O). The obtained binding energy of -0.34 eV is comparable to that of the NiPc COF monolayer (-0.37 eV), indicating NbPc COF monolayer is thermally stable. Furthermore, our calculated cleavage energy of 0.87 Jm$^{-2}$ is close to ~0.40 Jm$^{-2}$ of graphite or black phosphorus [34,35], suggesting that the NbPc COF monolayer can be possibly exfoliated from the corresponding bulk phase in real experiments.

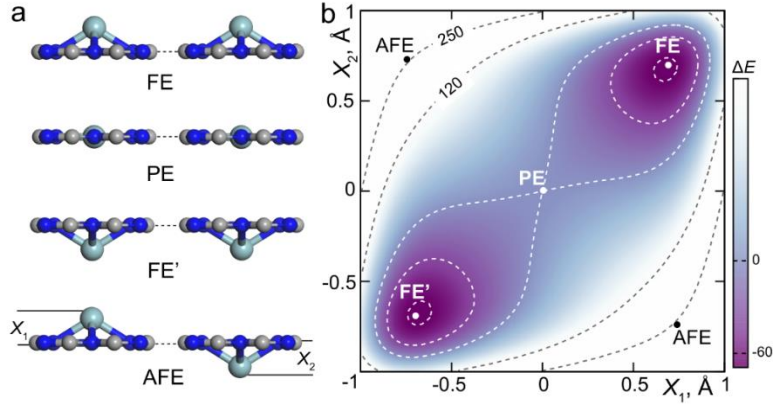

**Fig. 2.** (a) Sketch of Nb's displacement in FE, FE', AFE and PE phase of NbPc COF monolayer. $X_1$ and $X_2$ represent the Nb's and its nearest neighbors' displacements, respectively. FE, and FE' phase have opposite atomic displacements. (b) Free-energy contour plot as a function of the Nb's displacements, $X_1$ and $X_2$; FE, FE', AFE and PE are marked. Here, PE phase is taken as reference ($\Delta E$, meV).

## B. Ferroelectricity

To determine the most stable phase, the free-energy contour is calculated by scanning different nearest neighbor Nb displacements, marked as $X_1$ and $X_2$ in Fig. 2(a). Remarkably, NbPc COF monolayer has two equivalent minima configurations along the diagonal direction ($X_1 = X_2 = \pm 0.73$ Å), which are separated by the nonpolar PE state ($X_1 = X_2 = 0$ Å). Importantly, these two minima in the anharmonic potential surface exhibit opposite polarizations, labeled as FE and FE' in Fig. 2(b), serving as strong evidence for the existence of ferroelectricity in the NbPc COF monolayer. On the other hand, we found that the structure with opposite signs of $X_1$ and $X_2$ (AFE state) will become unfavorable in NbPc COF monolayer, as shown in Fig. 2(b). The strong covalent bonds between Nb and N ions may cause short-range dipole interactions to deviate significantly from their asymptotic form. These short-range interactions dominate the phase energetics. Hence, in the NbPc COF monolayer, the energy barrier for the transition from the PE to the FE phase is lower than that for the transition to the AFE phase, as illustrated in Fig. 2. Hence, we can naturally define the out-of-plane displacement of the Nb atom as $X_1 = X_2 = \pm 0.73$ Å.

We then proceed to estimate the magnitude of the ferroelectric polarization by employing the standard Berry phase calculations. The value of out-of-plane ferroelectric polarization can be calculated as $XZ^*$, where $Z^*$ indicates the out-of-plane Born effective charge of Nb. The Berry phase calculation gives a spontaneous polarization of $4.63 \times 10^{-12}$ C/m for 2D NbPc COF if PE structure is taken as reference. This value is much smaller than in-plane polarization due to out-of-plane depolarization effects, but it is found to be close to that predicted in 2D honeycomb binary buckled compounds [36]. Furthermore, the anharmonic double-well energy curve obtained by the NEB method in Fig. 3(b) is consistent with the ferroelectric $\varphi^6$ potential, reinforcing the ferroelectric nature of NbPc COF monolayer. The calculated ferroelectric depolarized barrier is about 67.2 meV/Nb, which is directly associated with the high ferroelectric Curie temperature, indicating the strong ferroelectric instability for 2D NbPc COF. To investigate the phase transition in ferroelectric 2D NbPc COF, the MC simulation was performed within Landau-Ginzburg theory [37]. The total energy of ferroelectric 2D NbPc COF is written as:

$$H = \sum_i \frac{A}{2} P_i^2 + \frac{B}{4} P_i^4 + \frac{C}{6} P_i^6 + \frac{D}{2} \sum_{\langle i,j \rangle} (P_i - P_j)^2, \quad (1)$$

where the first three terms correspond to the ferroelectric double-well potential. The last term represents the interaction between nearest neighbors $P_i$ and $P_j$, which can be understood as a discretized grid version of $(\Delta P)^2$, commonly known as the Ginzburg term. The parameters $A$, $B$, and $C$ in Eq. (1) are determined by fitting the double-well potential as shown in Fig. 3(b). The parameter $D$ is fitted by using the mean-field approach, where energy change in 2D NbPc COF is considered a square function of the displacement of Nb atoms relative to their neighbors [Fig. 3(c)]. The calculated values of $A$, $B$, $C$, and $D$ are listed in Table I.

**TABLE I.** Fitted to match direct DFT Data: $A$ (meVm$^2$C$^{-2}$), $B$ (meVm$^4$C$^{-4}$), $C$ (meVm$^6$C$^{-6}$), $D$ (meVm$^2$C$^{-2}$) parameters of Eq. (1). $E_p$ is energy barrier in unit of meV. $P_s$ is the spontaneous polarization in unit of $10^{-12}$ C/m. $T_C$ is the Curie temperature in unit of K.

| $A$ | $B$ | $C$ | $D$ | $E_p$ | $P_s$ | $T_C$ |
|---|---|---|---|---|---|---|
| -8.84 | -0.64 | 0.06 | 1.88 | 76.2 | 4.63 | 396(MC)/330(MD) |

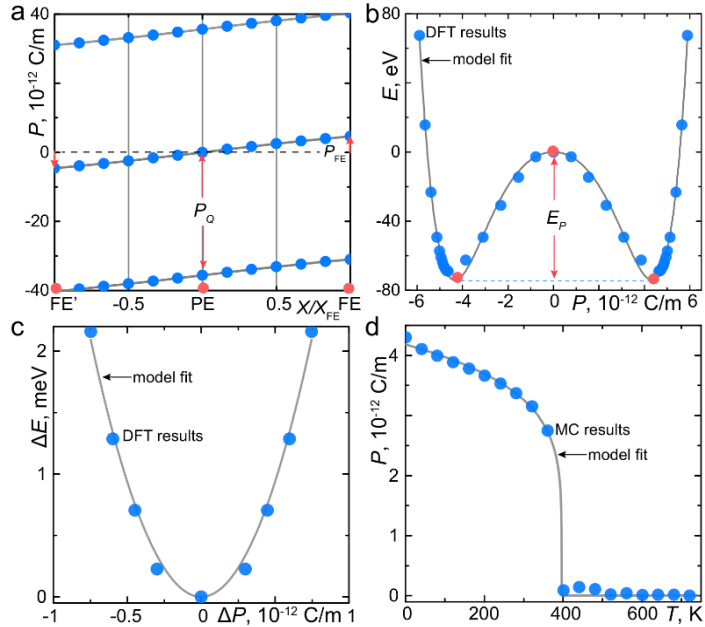

**Fig. 3.** Ferroelectricity of NbPc COF monolayer. (a) Polarization $P$ as a function of atomic displacement, $X$. $P_Q$ is polarization quantum. (b) Energy $E$ as a function of $P$ or $X$; repolarization barrier $E_p$ is marked. (c) Nearest-neighbor dipole-dipole interaction in 2D NbPc COF based on mean field theory. (d) FE

polarization *P* as a function of temperature *T*: MC data (dots) and fitted curves.

With the fitted parameter in Eq. (1), we can move to investigate the ferroelectric phase transition in the 2D NbPc COF using the MC method. The MC simulations give $T_C$ ~ 396K, which can be attributed to its sizable depolarized barrier. Furthermore, *ab initio* molecular dynamics (AIMD) simulations show the transition temperature is very close to MC results ($T_C$ ~ 330 K in Fig. S2) [33]. Due to its Curie transition point above room temperature, 2D NbPc COF shows attractive potential for practical applications. To further understand the universal critical phenomena, we adopt a fitting procedure that assumes a heuristic form for the temperature-dependent polarization *P(T)* [37],

$$P(T) = \begin{cases} \mu(T_C - T)^\delta, & T < T_C \\ 0, & T \geq T_C \end{cases}, \quad (2)$$

where $T_C$ is the Curie temperature, $\delta$ is the critical exponent, and $\mu$ is a constant [38]. The critical exponent $\delta$ is about 0.18, which is very close to the value (1/8) of 2D the Ising model [39]. This $\delta$ value suggests the ferroelectric behavior of 2D NbPc COF is not solely governed by short-range interactions, while long-range Coulomb coupling also contributed to stabilizing the ferroelectric polarization. Furthermore, the negative coefficient for the $P^4$ term in Eq. (1) indicates the first-order ferroelectric transition in the 2D NbPc COF, which is consistent with the temperature-dependent polarization obtained by MC simulation in Fig. 3(d).

## C. Magnetism

In the 2D NbPc COF, two central hydrogen atoms in the phthalocyanine are substituted with one Nb atom, leading to Nb ions with a nominal valence +2. Taking account of the valence configuration of $Nb^{2+}$ ($4d^3$), it is conceivable that the 2D NbPc COF could exhibit magnetic ordering if the $Nb^{2+}$ ions are in the high spin state. We then proceed to conduct the spin-polarized calculations for 2D NbPc COF. Here, the antiferromagnetic (AFM), ferromagnetic (FM) and non-magnetic (NM) ordering are employed to investigate the magnetic ground state of 2D NbPc COF in both PE and FE phases. With the Hubbard-type correlations for Nb's 4d orbitals, we systematically scan a range of $U_{eff}$ (Nb) from 0 to 3 eV (Table II for FE phase, Table SI for PE phase) [33]. When $U_{eff}$ (Nb) < 2 eV, the calculated magnetic ground states of both PE and FE 2D NbPc COF are AFM ordering. Moreover, the local magnetic moment of Nb reaches a convergence value of 2 $\mu_B$ in both PE and FE phases, indicating the presence of low spin states in the Nb $4d^3$ orbitals. Furthermore, upon comparing the energies of FE and PE phases in the AFM state, we found that the ferroelectric ground state in 2D NbPc COF is not influenced by the AFM ordering.

Interestingly, with the increase of $U_{eff}$, the magnetic ground state of the FE NbPc COF transitions from AFM to FM when $U_{eff} \geq 2$ eV. Unlike the 3d orbitals, exhibiting stronger electron exchange-correlation, the $U_{eff}$ for $Nb^{2+}$ ($4d^3$) should be smaller than that for $V^{2+}$ ($3d^3$), where typical $U_{eff}$ ($V^{2+}$) is around 3 eV [16,40]. Given the absence of experimental results for the NbPc COF monolayer, both $U_{eff}$ = 1 and 2 eV are reasonable choices for describing the correlation effects of Nb 4d orbitals. Actually, the similar $U_{eff}$ parameter for $Nb^{2+}$ ($U_{eff}$ = 1.5 eV) was also adopted in previous studies [41].

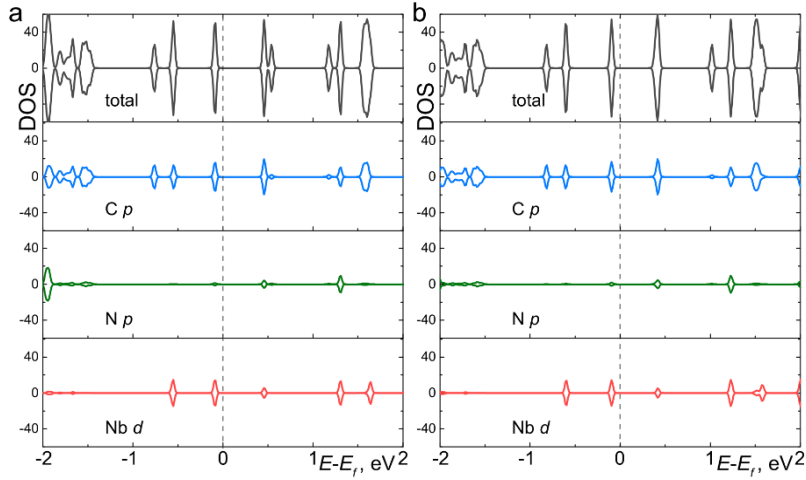

**Fig. 4.** Calculated density of states (DOS) of ground AFM 2D NbPc COF when $U_{eff}$ = 1 eV. Cs' p orbitals, Ns' p orbitals and Nbs' d orbitals are also projected to total DOS for PE(a) and FE(b).

The electronic structures of both the PE and FE phases of the 2D NbPc COF are shown in Fig. 4. With $U_{eff}$ = 1 eV, both phases exhibit semiconductor behavior with band gaps of 0.40 eV and 0.35 eV, respectively. As expected, the partial density of states (PDOS) show that states around Fermi level are predominantly attributed from the Nb's 4d orbitals. Additionally, the PDOS reveals the slight hybridization between the Nb 4d orbitals and the benzene moieties, which may serve as a bridge for magnetic coupling effects. When $U_{eff}$ = 2 eV, the electronic structure of FE phase shows ferromagnetic half-metal (Fig. S4) [33].

Despite the significant distance between neighboring Nb ions in 2D NbPc COF (23.2 Å), there is a possibility of magnetic coupling and even long-range ordering mediated by the organic linkage. This behavior is confirmed in previous dicopper (II) metallacyclophane with long distance (15 Å) [42], where magnetic exchange interactions are mediated through the double p-diphenylethynediamidate bridges. The Heisenberg spin model could provide a further understanding of the magnetic transition of local Nb's 4d orbitals, which expresses as:

$$H = -J \sum_{\langle i,j \rangle} S_i S_j + \sum_i A_z S_{iz}^2 , \qquad (3)$$

where the first item represents the exchange interaction between nearest-neighbor Nb ions, while the second term accounts for magnetic anisotropy. The coefficient $A_z$ corresponds to the magnetocrystalline anisotropy, and $S_z$ represents the spin component along the magnetic easy axis. With the spin magnitude normalized to |S|=1, the coefficients in Eq. (3) are calculated by considering FM, AFM, and NM orderings. With $U_{eff}$ = 1 eV, the obtained values of $J$ = -0.61 meV for FE phase indicate the AFM coupling between nearest-neighbor Nb ions. Moreover, the magnetic anisotropy calculations suggest that the spins tend to align along the $z$-direction. Now, with all the calculated parameters in Eq. (3), the AFM ground state is verified through unbiased MC simulations. The estimated Neel temperature ($T_N$) for FE phase is found to be approximately 5.6 K [Fig. 5(a)], which is lower compared to other 2D systems. This can be attributed to the larger magnetic distance and weaker magnetic coupling in 2D NbPc COF. With $U_{eff}$ = 2 eV, the obtained $J$ changes to 0.55 meV for FE phase. The estimated Curie temperature ($T_C$) for FE phase is approximately 9.8 K (Fig. S5) [33].

**TABLE II.** Calculated energy (unit: meV) of nonmagnetic (NM), ferromagnetic (FM), antiferromagnetic (AFM) states for ferroelectric (FE) phase with various Hubbard $U_{eff}$ (NM is taken as reference). The magnetic moment ($M$) of ground state is listed in unit of $\mu_B$. The Heisenberg exchange interaction $J$ and Magnetic anisotropy energy $A_z$ are expressed in unit of meV.

| $U_{eff}$ | NM | FM | AFM | $M$ | $J$ | $A_z$ |
|---|---|---|---|---|---|---|
| 0 | 0 | -807.6 | -810.1 | 1.70 | -0.63 | 0.21 |
| 1 | 0 | -665.5 | -667.9 | 1.50 | -0.61 | 0.25 |
| 2 | 0 | -1087.6 | -1085.4 | 1.92 | 0.55 | 3.32 |
| 3 | 0 | -978.8 | -969.9 | 1.85 | 2.23 | 0.07 |

**D. Magnetoelectric Coupling**

In conventional 2D type-I multiferroics, such as vdWs multiferroic heterostructures, the origins of magnetism and ferroelectricity stem from different sources, leading to much weaker ME coupling. However, in the case of 2D NbPc COF, the presence of a high-spin Nb $4d^3$ state contributes to the magnetism, while atomic Nb displacements generate the ferroelectric polarization. Both magnetism and ferroelectricity arise from the same origin, suggesting that 2D NbPc COF may exhibit a strong ME coupling. Considering the weaker magnetic coupling in 2D NbPc COF, the magnetic lower transition point is significantly below room temperature [Fig. 5(a)], while the ferroelectric polarization can have a stronger influence on its magnetism. Therefore, the discussion of magnetoelectricity in the 2D NbPc COF is limited around magnetic transition temperature.

To investigate the ME coupling, the magnetic exchange interaction $J$ is calculated as a function of normalized Nb displacements, as shown in Fig. 5(b). As expected, with $U_{eff}$ = 1 eV, the value of $J$ in the FE state (-0.61 meV) is approximately 762.5% stronger than that in the PE state (-0.08 meV/Nb), indicating the strong ME coupling in 2D NbPc COF. Interestingly, when $U_{eff}$ = 2 eV, the obtained $J$ is 0.55 meV for FE phase, while $J$ changes to -0.08 for PE phase (Table SI) [33]. These changes in $J$ from PE to FE phase (762.5% for $U_{eff}$ = 1 eV, -787.5% for $U_{eff}$= 2 eV) are much larger compared to bulk 3D multiferroic perovskites and other inorganic 2D systems, such as BiFeO$_3$ (3.8%), GdWN$_3$ (-47%) and 2D CrN (190.9%) [43–45]. This remarkable ME effect can be attributed to the changes in bond length and angle. The magnetic exchange interactions can be affected through the organic linkage [Fig. 3(b)]. Hence, any alterations in the bonds of the organic linkage induced by the displacement of Nb may impact the ME coupling. Interestingly, our MC snapshots [inserts of Fig. 5(b)] exhibit a tendency of disorder-order transition in the spin texture that is influenced by the Nb displacements, providing further evidence for the robust ME coupling. Therefore, the external electric field enables the manipulation of local magnetism within a specific temperature range.

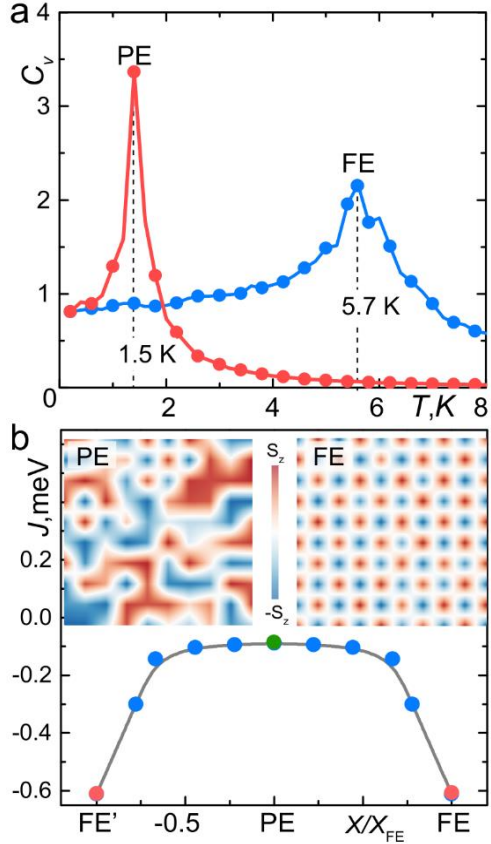

**Fig. 5.** MC results with $U_{eff}$ = 1 eV. (a) The magnetic heat capacity ($C_v$) dependent temperature ($T$) for PE and FE, respectively. (b) The magnetic exchange interaction $J$ as a function of normalized Nb displacements. Inserts: calculated spin textures for PE and FE phase under T=2K.

**CONCLUSIONS**

In summary, the coexistence of antiferromagnetic and ferroelectric order is predicted in NbPc COF monolayer. Our MC simulations suggest that the ferroelectric Curie temperature is estimated to be above room temperature. Despite the large distance between neighboring Nb ions, the organic linkages are found to play significant roles in the magnetic coupling and the stabilization of long-range magnetic ordering. Importantly, due to the same origins of magnetism and ferroelectricity, the 2D NbPc COF exhibits the strong ME coupling, which is two or three times stronger than that observed in other type-I multiferroics. The crossover between 2D materials and magnetic ferroelectrics presents an intriguing topic with fundamental implications and potential benefits for nanoscale devices.

## ACKNOWLEDGMENT

This work was supported by the National Natural Science Foundation of China (Grant No. 12404102), the Natural Science Foundation of the Jiangsu Province (Grant No. BK20230806), open research fund of Key Laboratory of Quantum Materials and Devices of Ministry of Education (Southeast University), and Southeast University Interdisciplinary Research Program for Young Scholars (Grant No. 2024FGC1008). Most calculations were done on Big Data Center of Southeast University.